\documentclass[letter,traditabstract,longauth]{aa} 
\usepackage{graphicx}
\usepackage{natbib}

\usepackage{txfonts}
\begin{document}
\newcommand{\Zsolar}{\mbox{$\,\rm Z_{\odot}$}}
\newcommand{\Msolar}{\mbox{$\,\rm M_{\odot}$}}
\newcommand{\Lsolar}{\mbox{$\,\rm L_{\odot}$}}
\newcommand{\xs}{$\chi^{2}$}
\newcommand{\dxs}{$\Delta\chi^{2}$}
\newcommand{\xsn}{$\chi^{2}_{\nu}$}
\newcommand{\ls}{{\tiny \( \stackrel{<}{\sim}\)}}
\newcommand{\gs}{{\tiny \( \stackrel{>}{\sim}\)}}
\newcommand{\asec}{$^{\prime\prime}$}
\newcommand{\amin}{$^{\prime}$}
\newcommand{\mstar}{\mbox{$M_{*}$}}
\newcommand{\hi}{H{\sc i}\ }
\newcommand{\hii}{H{\sc ii}\ }
\newcommand{\kms}{$\rm km~s^{-1}$}

   \title{Herschel-SPIRE observations of the disturbed galaxy NGC4438\thanks{{\it Herschel} is an ESA space observatory with science instruments provided 
   by European-led Principal Investigator consortia and with important participation from NASA.}}

   \author{	  L. Cortese\inst{1},
	  G. J. Bendo\inst{2},
	  A. Boselli\inst{3},
	  J. I. Davies\inst{1},
	  H. L. Gomez\inst{1},
	  M. Pohlen\inst{1},
	  R. Auld\inst{1},
	  M. Baes\inst{4},
	  J. J. Bock\inst{5},
	  M. Bradford\inst{5},
	  V. Buat\inst{3},
	  N. Castro-Rodriguez\inst{6},
	  P. Chanial\inst{7},
	  S. Charlot\inst{8},
	  L. Ciesla\inst{3},
	  D. L. Clements\inst{2},
          A. Cooray\inst{9},
	  D. Cormier\inst{7},
	  E. Dwek\inst{10},
	  S. A. Eales\inst{1},
	  D. Elbaz\inst{7},
	  M. Galametz\inst{7},
	  F. Galliano\inst{7},
	  W. K. Gear\inst{1},
          J. Glenn\inst{11},
	  M. Griffin\inst{1},
	  S. Hony\inst{7},
	  K. G. Isaak\inst{1,11},
	  L. R. Levenson\inst{5},
	  N. Lu\inst{5},
	  S. Madden\inst{7},
	  B. O'Halloran\inst{2},
	  K. Okumura\inst{7},
	  S. Oliver\inst{13},
	  M. J. Page\inst{14},
          P. Panuzzo\inst{7},
	  A. Papageorgiou\inst{1},
	  T. J. Parkin\inst{15},
	  I. Perez-Fournon\inst{6},
	  N. Rangwala\inst{11},
	  E. E. Rigby\inst{16},
	  H. Roussel\inst{8},
	  A. Rykala\inst{1},
	  N. Sacchi\inst{17},
	  M. Sauvage\inst{7},
	  B. Schulz\inst{18},
	  M. R. P. Schirm\inst{15},
	  M. W. L. Smith\inst{1},
	  L. Spinoglio\inst{17},
	  J. A. Stevens\inst{19},
	  S. Srinivasan\inst{8},
	  M. Symeonidis\inst{14},
	  M. Trichas\inst{2},
	  M. Vaccari\inst{20},
	  L. Vigroux\inst{7},
	  C. D. Wilson\inst{15},
	  H. Wozniak\inst{21},
	  G. S. Wright\inst{22},
	  W. W. Zeilinger\inst{23}
          }

\institute{	
	School of Physics and Astronomy, Cardiff University, Queens  
Buildings The Parade, Cardiff CF24 3AA, UK
	 \and
		Astrophysics Group, Imperial College, Blackett Laboratory, Prince  
Consort Road, London SW7 2AZ, UK
	 \and
		Laboratoire d'Astrophysique de Marseille, UMR6110 CNRS, 38 rue F.  
Joliot-Curie, F-13388 Marseille France
          \and
		Sterrenkundig Observatorium, Universiteit Gent, Krijgslaan 281 S9,  
B-9000 Gent, Belgium
	 \and
		Jet Propulsion Laboratory, Pasadena, CA 91109, United States;  
Department of Astronomy, California Institute of Technology, Pasadena,  
CA 91125, USA
	\and
		Instituto de Astrof\'isica de Canarias, v\'ia L\'actea S/N, E-38200 La  
Laguna, Spain
	\and
CEA, Laboratoire AIM, Irfu/SAp, Orme des Merisiers, F-91191
Gif-sur-Yvette, France
	\and
		Institut d'Astrophysique de Paris, UMR7095 CNRS, Universit\'e Pierre  
\& Marie Curie, 98 bis Boulevard Arago, F-75014 Paris, France
\and
Department of Physics \& Astronomy, University of California, Irvine,
CA 92697, USA 
              \and	
		Observational  Cosmology Lab, Code 665, NASA Goddard Space Flight   
Center Greenbelt, MD 20771, USA
	\and
		Department of Astrophysical and Planetary Sciences, CASA CB-389,  
University of Colorado, Boulder, CO 80309, USA
\and
ESA Astrophysics Missions Division, ESTEC, PO Box 299, 2200 AG
		Noordwijk, The Netherlands
	\and
		Astronomy Centre, Department of Physics and Astronomy, University of  
Sussex, UK
	\and
		Mullard Space Science Laboratory, University College London,  
Holmbury St Mary, Dorking, Surrey RH5 6NT, UK
	\and
		Dept. of Physics \& Astronomy, McMaster University, Hamilton,  
Ontario, L8S 4M1, Canada
	\and
		School of Physics \& Astronomy, University of Nottingham, University  
Park, Nottingham NG7 2RD, UK
	\and
		Istituto di Fisica dello Spazio Interplanetario, INAF, Via del Fosso  
del Cavaliere 100, I-00133 Roma, Italy
	\and
		Infrared Processing and Analysis Center, California Institute of  
Technology, Mail Code 100-22, 770 South Wilson Av, Pasadena, CA 91125,  
USA
	\and
		Centre for Astrophysics Research, Science and Technology Research  
Centre, University of Hertfordshire, College Lane, Herts AL10 9AB, UK
	\and
		University of Padova, Department of Astronomy, Vicolo Osservatorio  
3, I-35122 Padova, Italy
	\and
		Observatoire Astronomique de Strasbourg, UMR 7550 Universit\'e de  
Strasbourg - CNRS, 11, rue de l'Universit\'e, F-67000 Strasbourg
\and
UK Astronomy Technology Center, Royal Observatory Edinburgh, Edinburgh, EH9 3HJ, UK 
	\and
		Institut f\"ur Astronomie, Universit\"at Wien, T\"urkenschanzstr. 17,  
A-1180 Wien, Austria
}
   \date{Submitted to A\&A Herschel Special Issue}

 
  \abstract{We present {\it Herschel}-SPIRE observations of the perturbed galaxy NGC4438
in the Virgo cluster. These images reveal the presence of extra-planar dust 
up to $\sim$4-5 kpc away from the galaxy's disk. 
The dust closely follows the distribution of the stripped atomic and molecular hydrogen, 
supporting the idea that gas and dust are perturbed in a similar fashion by the 
cluster environment. Interestingly, the extra-planar dust lacks a warm temperature component when 
compared to the material still present in the disk, explaining why it was missed by previous 
far-infrared investigations.
Our study provides evidence for dust stripping in clusters of galaxies and 
illustrates the potential of {\it Herschel} data for our understanding of environmental 
effects on galaxy evolution.
}

   \keywords{Galaxies: evolution -- Galaxies: individual: NGC4438 -- Infrared: galaxies -- ISM: dust}

	\authorrunning{Cortese et al.}	
   \maketitle
%

\section{Introduction}
Clusters of galaxies are extremely hostile environments for 
star-forming galaxies. A plethora of observations and numerical simulations 
have revealed how gravitational and hydrodynamical interactions can affect 
cluster spirals (e.g., \citealp{review}): stars and gas can be stripped, 
galaxy morphologies can be changed and star formation activity can be enhanced and/or quenched. 
However, still very little is known about the effects of the environment on the dust content 
of cluster galaxies. Since the dust is mixed with the interstellar medium (ISM), 
the common expectation is that the 
environment should affect the dust in a similar fashion as the gas. 
However this hypothesis has still to be confirmed observationally.
The launch of {\it Herschel} \citep{pilbratt10} has opened a new era in the study of 
dust in galaxies. Thanks to its high spatial resolution and sensitivity to 
all dust components, {\it Herschel} should be able to determine whether dust is stripped from 
infalling cluster spirals and dispersed into the intra-cluster medium (ICM). 

\begin{figure*}
   \centering
   \includegraphics[width=15.5cm]{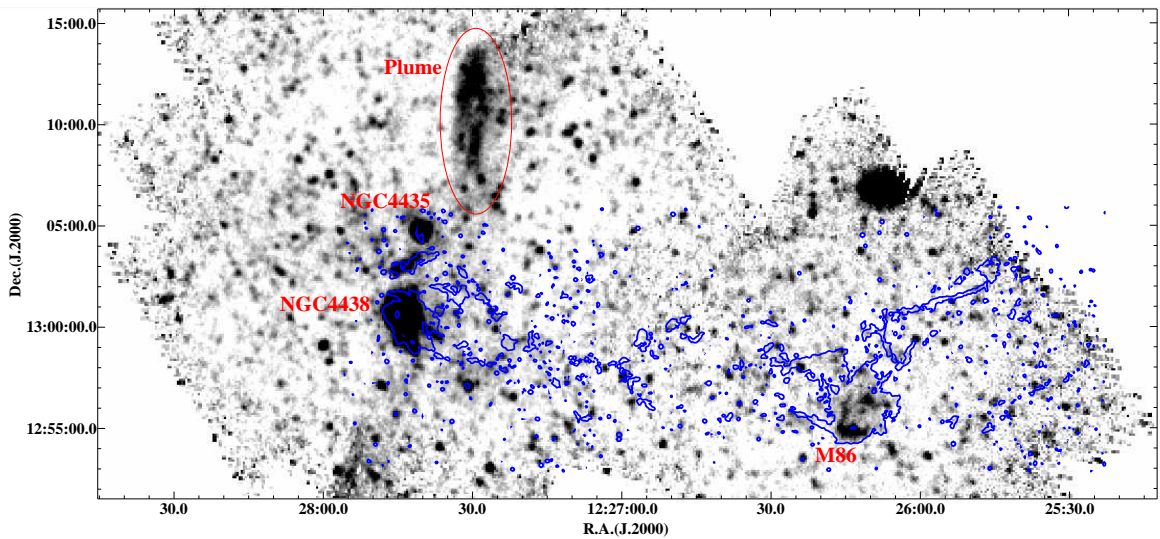}
      \caption{The NGC4438/M86 region as seen by SPIRE. This mosaic is obtained by combining the 250$\mu$m maps centered on NGC4438 and M86 \citep{gomez10}. 
      The blue contours show the extended H$\alpha$+[NII] filaments discovered by \cite{kenney08}.  
      Note that the intensity excess south of NGC4438 (near the edge of the frame) is just a residual of the baseline subtraction.}
         \label{spiremap}
   \end{figure*}

One of the most dramatic examples of the effects of the environment 
on nearby galaxies is represented by the disturbed galaxy NGC4438, in the Virgo cluster.
NGC 4438 is a highly H{\sc i}-deficient early-type spiral showing stellar tails \citep{kenney95,n4438} 
and extra-planar gas \citep{vollmer09}. 
Numerical simulations show that only a combination of ram-pressure stripping and tidal interactions 
is able to reproduce the disturbed morphology and kinematics of NGC4438 \citep{vollmer05}.
However, while \cite{combes88} and \cite{vollmer05} invoked a high-velocity ($\sim$800 \kms) gravitational interaction
 with the companion early-type galaxy NGC4435, \cite{kenney08} have 
revealed the presence of a series of H$\alpha$+[NII] filaments connecting NGC4438 to the giant elliptical M86 (see Fig.~\ref{spiremap}). 
These new observations favour a more complex scenario in which NGC4438 has recently ($\sim$100 Myr ago) 
interacted with M86 \citep{kenney08}.
Although the detailed history of NGC4438 is still unclear, its peculiar properties make it 
an ideal target to test the power of {\it Herschel} in unveiling the effects of the environment 
on the dust properties of cluster galaxies.
 
In this Letter we present {\it Herschel}-SPIRE \citep{griffin10} observations of NGC4438 obtained 
as part of the {\it Herschel} Science Demonstration (SD) Phase. The observations of the elliptical galaxy 
M86 are presented in a companion paper \citep{gomez10}. We 
assume for NGC4438 a distance of 17 Mpc \citep{gav99}, corresponding 
to a linear scale of 82 pc/\arcsec.

\section{Observations and data reduction}
The region around NGC4438 was observed by the {\it Herschel}-SPIRE instrument as part of the 
SD observations for the Herschel Reference Survey \citep{HRS}.
Eight pairs of cross-linked scan-map observations were carried out over 
an area of $\sim$12\arcmin$\times$12\arcmin~with a nominal 
scan speed of 30\arcsec/sec.
The data have been reduced following the procedure described 
in \cite{pohlen10} and \cite{bendo10}. However, since in this case 
the median baseline of the whole time line (prior to the de-striper) left some residual 
large-scale gradients (see \citealp{pohlen10}), 
we preferred an alternative baseline subtraction 
on a scan by scan basis.
A robust linear fit with outlier rejection was applied to the 
first and last fifty sample points of the time line for each bolometer, 
thus avoiding the galaxies and the extended diffuse emission present 
across the field.
The SPIRE astronomical calibration methods and accuracy are outlined in \cite{swinyard10}.
Since at the time of the data reduction the SPIRE pipeline used a preliminary flux calibration, 
we followed the recommendation of the SPIRE Instrument Control Center and multiplied the flux densities 
by 1.02, 1.05, and 0.94 at 250, 350, and 500 $\mu$m, 
respectively\footnote{See also http://herschel.esac.esa.int/SDP$\_$wkshops/presentations/
IR/3$\_$Griffin$\_$SPIRE$\_$SDP2009.pdf.}.
The images have flux calibration uncertainties of 15\% and 
the rms are $\sim$5.5, 5.8, 6.9 mJy/beam at 250, 350 and 500 $\mu$m, respectively.
We note that in all three bands the noise is dominated by confusion.
The astrometric uncertainty is $\sim$2\arcsec~and the full widths at half maximum 
of the SPIRE beams are 18.1\arcsec, 25.2\arcsec, 36.9\arcsec~at 250, 350 and 500 $\mu$m, respectively.

\section{Results}
In Fig.~\ref{spiremap} we show a mosaic of the 250 $\mu$m SPIRE maps 
of NGC4438 and M86 \citep{gomez10}.
In addition to NGC4438, also the companion S0 galaxy NGC4435 is 
clearly detected in all the three SPIRE bands. This 
is not surprising since \cite{panuzzo07b} showed that this galaxy 
has just experienced a burst of star formation and it is 
confirmed by the fact that the SPIRE flux density 
ratios ($f(250)/f(350)\sim$2.9 and $f(350)/f(500)\sim$2.6) are consistent 
with the values observed in star-forming galaxies \citep{boselli10}. 

More intriguing is the presence of a few bright (i.e., peak $f(250)\geq$0.05 Jy/beam) 
extended sources possibly associated with NGC4438.
The most remarkable feature is the extended emission to the N-NW of NGC4435 
(Plume in Fig.~\ref{spiremap}). This plume is approximately 8\arcmin~long 
and 3\arcmin~wide and it is detected (at least in part) in all the three SPIRE bands. 
Although its origin (i.e., Galactic or extragalactic) is still debated, 
\cite{cortese10} have recently shown that the properties of this plume are more consistent 
with Galactic cirrus than tidal debris at the distance of the Virgo cluster.

Two additional elongated structures 
(E1 and  E2 in Fig.~\ref{collage}) are detected within the optical radius of NGC4438. 
E1 appears as a tail extending from the bulge of NGC4438. 
However a careful comparison of the SPIRE data with {\it Spitzer} images 
reveals that it is just the result of blending of point sources visible 
at 3.6 and 24 $\mu$m, possibly background galaxies. 
This seems also supported by the lack of any counterpart in 
UV, \hi or H$\alpha$+[NII]. 
Less clear is the origin of E2, which is composed of three different 
knots. The southern and central knots coincide with an \hi cloud discovered by \cite{hota07} 
and the whole feature apparently follows one of the H$\alpha$+[NII] filaments 
pointing towards M86 \citep{kenney08}, 
suggesting that it might be associated with NGC4438 (see Fig.~\ref{collage}). 
If so, this would represent a unique example of stripped intra-cluster dust. 
However, caution is required before interpreting E2 as a dust stream 
in Virgo. Firstly, the southern knot is offset
($\sim$15\arcsec~north) from the H$\alpha$+[NII] stream, thus it is not clear 
whether the two features are related. 
Secondly, {\it Spitzer} 3.6 and 24 $\mu$m point sources are found within $\sim$6-9\arcsec~from each of the three knots, 
thus we cannot completely exclude the possibility that E2 is just the 
result of blending of background sources, like E1. 
So, although E2 is a very intriguing system, only future investigations will reveal whether it is 
associated with NGC4438. We note that, in this case, 
E2 would represent the only case of submillimetre (submm) emission associated with the intra-cluster 
H$\alpha$+[NII] streams connecting M86 and NGC4438 (see also \citealp{gomez10}).

Stronger evidence supporting dust stripping is found 
in the main body of NGC4438.
In Fig.~\ref{collage} we compare the distribution of the cold dust as 
revealed by {\it Herschel} with those of stars, hot/warm dust and warm ionized\footnote{We note that 
the H$\alpha$+[NII] in NGC4438 and the stripped filaments is likely not tracing star formation but gas 
cooling after shocks \citep{kenney95,machacek04,vollmer09}.}, cold atomic and molecular hydrogen.
Interestingly, despite the presence of a young stellar population, no submm emission 
is detected in the northern and southern tidal tails.
Although the 250-500 $\mu$m emission is mainly concentrated within the central $\sim$4.5 kpc of the galaxy, 
extended emission is visible on the west side of NGC4438. Particularly remarkable are the two 
bright regions highlighted in Fig.~\ref{collage}: a tail (A) extending to the south up to $\sim$9 kpc from 
the center of NGC4438, and an elongated structure (B) $\sim$4.5 kpc from the plane of the disk and 
almost completely detached from the main body of NGC4438. 
  \begin{figure*}
   \centering
   \includegraphics[width=17cm]{14547fg2.epsi}
      \caption{A panchromatic view of NGC4438. Each column shows the distribution 
      of a different baryonic component. First column: stars (GALEX UV, SDSS gri, 2MASS JHK).
      Second column: cold dust (SPIRE 250, 350, 500 $\mu$m). Third column: warm dust ({\it Spitzer} 8, 24, 70 $\mu$m). 
      Fourth column: cold atomic (H{\sc i}, \citealp{hota07}), molecular (CO, \citealp{vollmer05}) and warm ionized 
      (H$\alpha$+[NII], \citealp{kenney08}) hydrogen.
      The contours show the 250 $\mu$m emission. Contours levels are 0.03, 0.04, 0.07, 0.11, 
      0.15, 0.19 Jy beam$^{-1}$. The features discussed in \S~3 are highlighted in the 250 $\mu$m map.}
         \label{collage}
   \end{figure*}

The tail A is detected at 8, 24, 70 $\mu$m and in H{\sc i}. 
It extends up the southern stellar tail of 
NGC4438 and is composed of at least two disjoint bright knots. 
Although it follows very closely the southernmost H$\alpha$+[NII] filament in NGC4438, 
a careful comparison between the 8, 24 $\mu$m and H$\alpha$+[NII] images reveals that 
two features do not spatially coincide. The fact that the \hi in correspondence of tail A has 
a recessional velocity ($-$40$<V<+$20 \kms, \citealp{hota07}) 
significantly lower than the ionized gas ($V\sim -$85 \kms, \citealp{chemin05}) suggests 
that the two components might just be projected along the same line-of-sight.
Although the presence of submm emission up to the stellar tidal tail is consistent with a 
dust stripping scenario, it remains uncertain whether tail A is just part of what 
is left of the disk of NGC4438 or a stripped dust tail.

In comparison, the formation history of knot B appears a little bit clearer.
This feature coincides with the extra-planar dust-lane visible in optical images and it clearly follows 
the distribution  of the extra-planar atomic \citep{hota07} and molecular \citep{vollmer05} hydrogen 
detected in this region. While the extra-planar CO(1-0) emission 
is mainly segregated around knot B, the \hi extends further south following closely the 
diffuse 250 $\mu$m emission. 
Although low surface brightness H$\alpha$+[NII] emission is observed in correspondence of the diffuse 250 $\mu$m 
to the west of NGC4438, only a single \hii region (slightly offset from the submm emission peak) is observed in knot B.
Previous investigations have shown that the disturbed morphology and kinematics of the gas in this 
region is consistent with a stripping scenario (\citealp{kenney95,combes88,vollmer05}).
So, it is likely that we are directly witnessing dust in the process of being removed from the disk of NGC4438.  
Contrary to tail A, knot B is not entirely detected by {\it Spitzer}: 
only the single \hii region and low surface brightness emission are visible at 8 and 24 $\mu$m.
The integrated $f(350)/f(24)$ flux density ratio of knot B ($\sim$47) is a 
factor $\sim$4 higher than the value observed in the main body of NGC4438, confirming 
that this feature is missing a warm dust component and it is not associated with active 
star formation. This is additionally supported by the {\it reddening} of the 
$f(250)/f(350)$ ratio (from $\sim$2.7 to $\sim$2.1) when moving from the center of NGC4438 to knot B.

\section{Discussion and conclusions}
The SPIRE data alone are not sufficient to determine whether the extra-planar dust 
is in the process of being removed from NGC4438 or, for example, falling back onto 
the disk.
However, by combining multiwavelength observations with detailed numerical simulations, 
\cite{vollmer05,vollmer09} have shown that the distribution and kinematics of the 
different components of the ISM in NGC4438 can only be reproduced via a combination 
of tidal interaction and ram-pressure stripping.
Although the details of the gravitational interaction are still unclear 
(i.e., M86 and/or NGC4435, \citealp{vollmer09b}), it appears that strong 
on-going ram-pressure (in addition to tidal forces) is necessary 
to reproduce the properties of the extra-planar gas component in knot B.
In this case, whatever the exact mechanisms affecting NGC4438, it is 
likely that at least part of the dust in the west side of the galaxy is in the process 
of being removed from the disk thus providing evidence of dust stripping 
by environmental effects. If completely removed, the stripped dust will be dispersed 
in the ICM contributing to its metal enrichment.

Since in galaxies the dust is associated with the gaseous component of the ISM, it is 
generally expected that when the gas is stripped part of the dust can be removed as well. 
What is still unclear is how much dust follows the fate 
of the stripped hydrogen. Unfortunately, the SPIRE fluxes alone are not sufficient 
to accurately quantify the amount and temperature of the dust in knot B. 
Nevertheless, we can at least try to estimate the dust mass by using the fluxes 
at 250 and 350 $\mu$m, where knot B is clearly resolved, and 
assuming $M_{dust} = (f_{\nu} D^{2})/[\kappa_{\nu} B(\nu, T)]$, where $f_{\nu}$ is the flux 
density ($\sim$0.98 and $\sim$0.46 Jy at 250 and 350 $\mu$m, respectively), 
$D$ is the distance and $\kappa_{\nu}$ is the absorption cross section per mass of dust.
We adopt $\kappa_{\nu}$=4 and 1.9 cm$^2$ g$^{-1}$  at 250 and 350 $\mu$m 
respectively \citep{draine03ara}.
The total mass of dust in knot B is in the range $\sim$2$\times$10$^{6}$-2$\times$10$^{7}$ M$_{\odot}$ 
for a dust temperature between 10 and 20 K.
In order to determine the total gas mass in the extra-planar cloud, we combined 
the estimates of M(H$_{2}$)$\sim$4.7$\times$10$^{8}$ M$_{\odot}$ (including helium) and 
M(H{\sc i}) $\sim$1.5$\times$10$^{8}$ M$_{\odot}$ obtained by \cite{vollmer05} and 
\cite{hota07}, respectively.
The total gas-to-dust ratio of knot B is in the range $\sim$30-300, not significantly 
different from the values observed in nearby galaxies (e.g., \citealp{draine07,galliano08}) and across the 
disk of star-forming spirals (e.g., \citealp{mateos09,bendo10b,pohlen10}).
This seems to support a scenario in which a cloud of gas does not lose 
a significant amount of its heavy elements when removed from the disk. 
It is in fact quite remarkable how well the cold dust follows the spatial 
distribution of the cold hydrogen across the whole galaxy, even in such a highly 
perturbed system like NGC4438.      

Interestingly, the high degree of spatial correlation between cold dust and cold hydrogen 
in NGC4438 is different from what is observed by {\it Herschel} in M86 \citep{gomez10}. 
In M86 the dust is mainly associated with the H$\alpha$+[NII] emitting gas and 
only mildly correlated with the cold atomic component. 
This is intriguing and might suggest different dust properties (e.g., stripping mechanism, 
heating source, etc.) in the two galaxies.

Finally, we note that it is unclear whether the extra-planar dust is missing a hot 
component just because the interstellar radiation field outside the plane is not strong 
enough to keep it hot, or if the cloud was already dominated by a cold dust component 
while in the disk. Although the $f(250)/f(350)$ flux ratio 
is lower than the value observed in the central part of NGC4438, it is not significantly 
different from what is observed in the outer parts of the disk of 
M81 \citep{bendo10}, M99 and M100 \citep{pohlen10}.

In summary, in this paper we provide evidence for dust stripping 
by environmental effects in NGC4438.
The high spatial resolution and sensitivity to cold dust of the SPIRE camera allowed us to 
discover an extra-planar cold dust component completely missed 
by previous far-infrared surveys.
The strong spatial correlation between cold dust, atomic and molecular hydrogen suggests that 
when the gas is removed also the dust is pulled out the galactic disk. 
These results provide interesting insights into the evolution of an extreme case 
among perturbed cluster galaxies.
Once combined with the discovery of truncated dust disks in H{\sc i}-deficient 
Virgo cluster spirals \citep{cortese10b}, our analysis clearly highlights  
the great potential of {\it Herschel} for our understanding of the 
effects of the cluster environment on the dust properties of galaxies.


\begin{acknowledgements}
We thank Ananda Hota, Jeff Kenney and Bernd Vollmer for providing us with an electronic 
version of their data.
SPIRE has been developed by a consortium of institutes led by 
Cardiff University (UK) and including Univ. Lethbridge (Canada); 
NAOC (China); CEA, LAM (France); IFSI, Univ. Padua (Italy); IAC (Spain); 
Stockholm Observatory (Sweden); Imperial College London, 
RAL, UCL-MSSL, UKATC, Univ. Sussex (UK); and Caltech, JPL, NHSC, 
Univ. Colorado (USA). This development has been supported by 
national funding agencies: CSA (Canada); NAOC (China); CEA, 
CNES, CNRS (France); ASI (Italy); MCINN (Spain); Stockholm 
Observatory (Sweden); STFC (UK); and NASA (USA).

\end{acknowledgements}

\end{document}